\documentclass{optica-article}

\journal{opticajournal} 

\articletype{Research Article}

\usepackage{lineno}
\usepackage{xcolor}
\usepackage{hyperref}
\usepackage{comment}
\usepackage{jabbrv}
\usepackage{graphicx}
\usepackage{float}

\begin{document}
\title{Principal Component Analysis in Application to Brillouin Microscopy Data}

\author{Hadi Mahmodi\authormark{1\dag}, Christopher G. Poulton\authormark{1 \dag}, Mathew N. Lesley\authormark{2},  Glenn Oldham\authormark{3}, Hui Xin Ong\authormark{2}, Steven J. Langford\authormark{1}, and Irina V. Kabakova\authormark{1*} }

\authormark{1} School of Mathematical and Physical Sciences, University of Technology Sydney, NSW, Ultimo, Australia

\authormark{2} Respiratory Technology, Woolcock Institute of Medical Research, NSW, Glebe, Australia 

\authormark{3} Swinburne University of Technology, Melbourne, Victoria, Australia

\authormark{*} irina.kabakova@uts.edu.au 

\authormark{\dag} equal contributions


\begin{abstract*} 
Brillouin microscopy has recently emerged as a new bio-imaging modality that provides information on the micromechanical properties of biological materials, cells and tissues. The data collected in a typical Brillouin microscopy experiment represents the high-dimensional set of spectral information. Its analysis requires non-trivial approaches due to subtlety in spectral variations as well as spatial and spectral overlaps of measured features. This article offers a guide to the application of Principal Component Analysis (PCA) for processing Brillouin imaging data. Being unsupervised multivariate analysis, PCA is well-suited to tackle processing of complex Brillouin spectra from heterogeneous biological samples with minimal {\em a priori} information requirements. We point out the importance of data pre-processing steps in order to improve outcomes of PCA. We also present a strategy where PCA combined with k-means clustering method can provide a working solution to data reconstruction and deeper insights into sample composition, structure and mechanics.    
\end{abstract*}

\section{Introduction}

Brillouin microscopy (BM) is a type of spectroscopic imaging where image contrast relies on the variation of micro-mechanical properties in matter \cite{Palombo_review,Poon_2021,Yakovlev:review}. These properties are obtained by direct detection of the speed and attenuation of hypersound waves. As a technology, BM has received a considerable attention in recent years due to advances in mechanobiology and the growing demand for label-free mechanical characterisation of biological materials, tissues and cells in 3D and at the spatial scales relevant to cellular and subcellular processes \cite{Wu:19,Polonchuk_2021,MAHMODI}. 

The data collected in a typical Brillouin imaging experiment consists of a few thousand individual spectra, each representing up to a thousand of points across the frequency range of interest. Therefore, high-dimensionality is one of the common challenges in analysis of Brillouin imaging data \cite{Xiang2021}. Each Brillouin spectrum demonstrates a set of peaks associated with inelastic scattering of light by hypersound waves inside the material. Solid state and glassy materials have relatively narrow peaks positioned sufficiently far apart, thus line-fitting methods can be straightforwardly applied to localise the peak's position and its full-width at half maximum - the two quantities needed to assess the material's mechanical properties.  

In biological matter, Brillouin signals exhibit broad bandwidth and their spectral position is quite close to the Brillouin frequency of water (owing to high hydration content of biomaterials, cells and tissues). Additionally, biological tissue and cells are heterogeneous across the imaging volumes traditionally employed in BM (a few cubic microns), leading to spectrally overlapping features. Therefore, another common challenge of Brillouin data analysis is associated with the subtle spectral variations across the data set accompanied by an overlap in spectral signatures of heterogeneous material components. 

These reasons make the analysis of Brillouin data collected from biological samples non-trivial, with simple line-fitting routines producing unsatisfactory results and often taking long processing time, not compatible with real-time imaging \cite{Xiang2021}. On the other hand, the methods common within the hyperspectral imaging community, in particular multivariate techniques such as Principal Component Analysis (PCA), can provide improvements in both the data processing speed and accuracy, thus supporting the translation of BM to medical diagnostics and routine clinical use \cite{RANDLEMAN2023}.

\begin{figure}[htbp]
\centering\includegraphics[width=11cm]{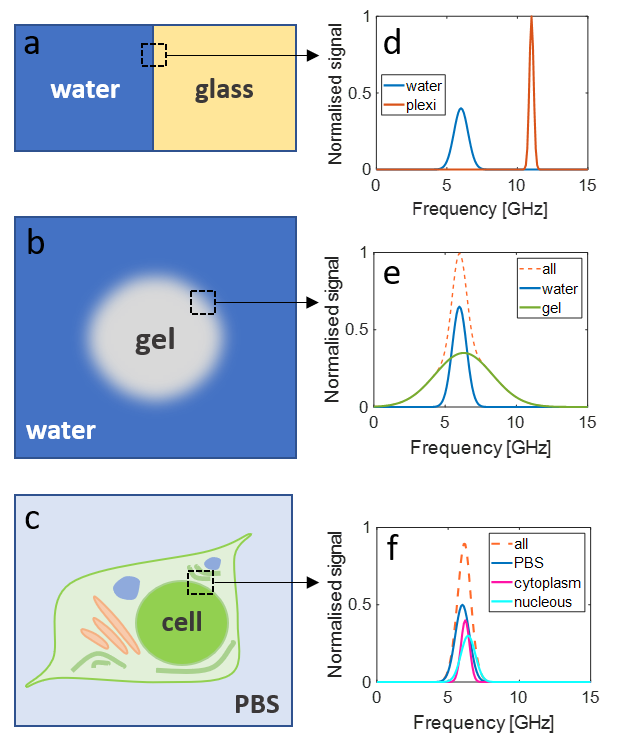}
\caption{Schematic depiction of the three sample types: (a) water and plexiglass interface, (b) hydrogel spheroid in water and (c) a eukaryotic cell. The black box indicates the region where a few material components within each sample are present. The associated Brillouin spectra for anti-Stokes peaks are schematically shown in (d), (e) and (f), respectively.}
\label{fig:fig1_samples}
\end{figure}

As a simple, non-parametric method for extracting useful information from confusing data, PCA is widely employed in various types of analysis, spanning from neuroscience to computer graphics \cite{Shlens_tutorial}. This approach enables the reduction of a complex data set to a lower dimension, unveiling the sometimes hidden dynamics that frequently underlies it. In fact, PCA is already routinely applied for analysis of hyperspectral data measured in standard Raman scattering experiments \cite{He_Raman} and coherent anti-Stokes Raman imaging \cite{Masia2013}. Within the Brillouin microscopy community, unsupervised multivariate techniques have so far been under-utilised, with only a few reports published to date \cite{Xiang2021, Palombo_Analyst2018,Cardinali2023}. It has been shown that the scores of principal components can be used as  markers for classification and sorting pathological samples from healthy tissues \cite{Cardinali2023}. Notably, the highest score principal component (PC1) has been deemed unusable for the algorithm's training purpose as it was found to represent variability in the signal intensity, which may be affected by the laser intensity noise and scattering volume fluctuations and thus, it is a less reliable characteristic of the sample's properties \cite{Cardinali2023}. Additionally, Xiang {\it et al.} have emphasized the difficulty in interpretation of PCA results applied to Brillouin scattering data that are associated with: i) coarse resolution of Brillouin imaging and consequent spectral mixing of multiple components within a single imaging voxel and ii) the method's sensitivity to spurious features such as the laser intensity fluctuations and its frequency drift \cite{Xiang2021}. Overall, PCA was found to be a sub-optimal method for spectral unmixing and signal processing of Brillouin imaging data compared to other supervised and unsupervised techniques \cite{Xiang2021}.       

In this article, we demonstrate that PCA is a simple and valuable method for understanding complex Brillouin scattering data collected from heterogeneous biological samples. By gradually increasing the sample complexity - from a water-plexiglass interface, to hydrogel spheroid and, finally, to a single cell (see Figure 1) - we are able to explain most features of the PCA functions and assess the method's resilience against possible problems in data quality,  such as signal-to-noise ratio and intensity fluctuations. We give step-by-step guidance for data pre-processing techniques that we believe are necessary to improve the outcomes of PCA method. Finally, we propose a new scheme in which PCA is combined with k-means clustering to enable spectral reconstruction and unmixing of Brillouin scattering data.

\section{Methods}
\subsection{Sample choice}
Three types of samples were used to carry out our study: 1) a plexiglass immersed in DI water, 2) a hydrogel spheroid in water and 3) human fibroblast cells. The choice of these three samples is motivated by the increasing level of complexity from sample 1 to 3 (see Figure 1 a-c). For example, sample 1 has two well-defined components (water and plexiglass) with distinctly different Brillouin frequencies 
($\nu_{\rm W}=\Omega_{\rm W}/2 \pi\approx$5.7~GHz and 
$\nu_{\rm P}=\Omega_{\rm P} / 2 \pi\approx$ 11~GHz) as schematically illustrated in Figure 1d. The hydrogel spheroid immersed in water undergoes a swelling process that results in a non-uniform distribution of the mechanical properties across it. Additionally, the Brillouin frequency of hydrogel ($\nu_{\rm H} = \Omega_{\rm H}/2\pi\approx$6~GHz) is quite close to that of surrounding water, resulting in a spectral line overlap (Figure 1e). Finally, a cell exhibits many intracellular components, each characterised by its frequency shift and linewidth, but overall closely spaced, leading to a complex asymmetric lineshape of the Brillouin scattered light (see Figure 1f). Such a sample selection helps us understand the features of PCA method applied first to data collected from a simple sample, and then translate this knowledge to more complex scenarios.      

\subsection{Sample preparation}
\subsubsection{Hydrogel fabrication}
2-Hydroxyethyl cellulose (average MW ~380,000) was purchased from Sigma Aldrich. The HEC hydrogel was created by adding water to the HEC polymer (20$\%$~w/v). The mixture was thoroughly stirred to create a homogeneous solution and left to set. When the gel was still liquid, it was poured into the desired disk mould to complete its formation. From there, the hydrogel was moved to a sealed container for storage to prevent dehydration. On the day of measurement, the hydrogel disk was immersed in water and left to swell for 1 hours. 

\subsubsection{Human fibroblast cells}
A healthy human lung fibroblast cell line, MRC-5 (ATCC CCL-171) was purchased from American Type Cell Culture Collection (ATCC) and incubated at 37 °C with 5$\%$ CO$_2$. The cell line was cultured in Modified Eagle Medium (Gibco) supplemented with 10$\%$ (v/v) heat-inactivated foetal bovine serum (Invitrogen), 7.5$\%$ (v/v) sodium bicarbonate (Gibco), 1mM sodium pyruvate (Gibco) and 1$\%$ (v/v) non-essential amino acids (Sigma-Aldrich). MRC-5 cells were seeded in 6-well plates (Corning Costar) coated with fibronectin (2 $\mu$g/cm$^2$; Sigma-Aldrich) at a density of 4 × 104 cells/cm$^2$. The plates were washed with Phosphate Buffered Saline (PBS) to remove unadhered cells 72 hours after seeding and fixed for 10 min at room temperature using 4$\%$ (v/v) paraformaldehyde/PBS. The MRC-5 cells were washed again then treated with PBS supplemented with 1$\%$ (v/v) Antibiotic-Antimycotic solution (Sigma-Aldrich). 

\subsection{Measurement system}
The system employed for collecting spontaneous Brillouin scattering spectra consisted of a confocal microscope integrated with a specialized Brillouin spectrometer, utilizing a 6-pass tandem scanning Fabry-Perot interferometer (TFP1, TableStable Ltd.). Sample illumination was achieved using a continuous frequency laser (660 nm, 120 mW, Torus, Laser Quantum) coupled to the confocal microscope (CM1, TableStable Ltd.) and focused onto the sample using a microscope objective (20X Mitutoyo Plan Apo infinity corrected objective, NA = 0.42, WD = 20 mm, and 60X Nikon CFI APO NIR Objective, 1.0 NA, 2.8 mm WD). Subsequently, the light scattered in the backward direction was collected by the same objective lens and directed to the 6-pass scanning Fabry-Perot interferometer for analysis. 
The spectral resolution of the Brillouin microscopy system is determined by the distance between the mirrors of the Fabry-Perot scanning interferometer (5 mm) and the number of acquisition channels (512), approximately equaling 276 MHz. The spectral extinction ratio of Fabry–Perot interferometers exceeds $10^{10}$ \cite{sandercock1976simple}. Spatial resolution is primarily influenced by the objective lens, potential aberrations within the confocal microscope, and spectrometer properties (input and output apertures). Considering these factors, our system's spatial resolution is estimated to be approximately 2 $\mu$m × 2 $\mu$m × 100 $\mu$m and 0.5 $\mu$m × 0.5 $\mu$m × 10 $\mu$m in the X-Y-Z directions for 20X and 60X objective lenses, respectively. The Brillouin data collection utilized in-house software for two-dimensional (2D) scans within a sample plane, offering an acquisition time of 1-20 s per point depending on the sample transparency. To prevent sample damage from incident radiation, laser power was maintained below 20 mW. Raw spectra of Brillouin scattered light, containing Rayleigh and Brillouin peaks (Stokes and anti-Stokes), were fitted using the damped harmonic oscillator (DHO) model, where the peak positions determine the BFS.

\subsection{Data pre-processing}

\subsubsection{Data signal-to-noise-ratio affects PCA outcomes}
The signal-to-noise ratio (SNR) of Brillouin data depends on many factors, but can be controlled via reducing or increasing the time interval over which the signal is acquired at every spatial location within the sample (the acquisition time). In most experiments there exists a trade-off between selecting a suitable acquisition time (to achieve a sufficient SNR) and keeping scan times for the entire sample to a reasonable duration. We have observed that the percentages of each principal component found in the data set is not fixed, but depends on the SNR (see correlation between the percentage of data explained by the first principal component, PC1, as a function of SNR for DI water measurements shown in Fig. S1). Thus, it might be necessary to increase data acquisition time in order to improve outcomes of data analysis performed by PCA.

\subsubsection{Data normalisation and spectral drift correction}
In BM experiments there are a few measurement artifacts that can modify Brillouin spectra and which must be corrected to avoid the PCA algorithm identifying these as sample' characteristic features. The first effect is the variation in the absolute intensity of the spectrum across the sample, which can
occur due to laser fluctuations over the measurement time and the presence of reflective/scattering interfaces along the optical path. We corrected for this 
intensity drift by normalising the spectral amplitudes (Fig. S2). Specifically, we scaled each measured spectrum linearly to lie between the values of 0 and 1 over the frequency range of interest (i.e. excluding the central laser peak). 

The second important extraneous effect is spectral drift, which is a common feature in Brillouin scattering spectroscopy and microscopy that originates from small changes in the laser wavelength over the duration of the experiment or long-term drifts in the optical system. Such drifts, if undetected and corrected, can result in errors for the parameters of interest, namely the Brillouin frequency shift, and thus need to be accounted for. To correct for spectral drift we recentre each spectrum to the average frequency value corresponding to the Stokes and anti-Stokes peaks.

\section{Analysis of Brillouin data from a water/plexiglass sample}
To demonstrate the different steps of the analysis, we first consider Brillouin spectra from a sample consisting of a plexiglass layer immersed in water (Figs.~1(a) and 2(a)). The measured region consisted of a 2D rectangle of dimensions 16.5~mm x 15.5~mm, with a step size of 0.5~mm, chosen across the interface between water and plexiglass. The approximate thickness of the water layer was 1.5~cm, and that of the plexiglass layer was 1cm. Each spectrum was taken with a  measurement time of 10~s. 

\subsection{PCA analysis of Brillouin spectral data}
\begin{figure}[H]
\centering\includegraphics[width=\textwidth]{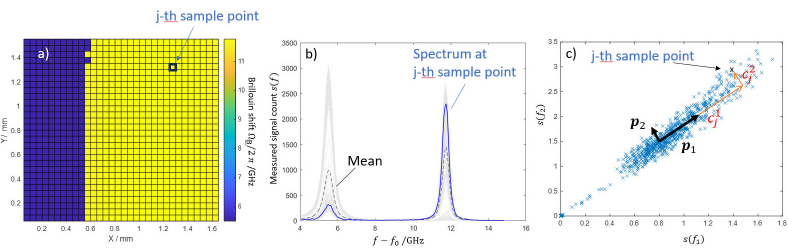}
\caption{Brillouin spectral data collected from a sample containing plexiglass and water. a) Map of the Brillouin frequency shift across the sample, as determined by fitting the data with DHO model. b) Non-normalised spectral data as a function of frequency for all points in the domain, showing the mean (black dashed) and spectrum for an arbitrary single point within the sample (blue solid). The peaks for water (near 5.7 GHz) and plexiglass (near 11.8 GHz) are clearly visible. c) Illustration of the correlation of spectral strength for two frequency points. The two-dimensional projections of the principal components are shown; each measured data point can be represented as a combination of the principal components.}
\label{fig:plexiglass_spectrum}
\end{figure}

Fig.~\ref{fig:plexiglass_spectrum}(a) shows the 
2D sample region, for which each pixel in this 2D map represents the Brillouin frequency shift at specific spatial location, computed using a Damped Harmonic Oscillator model to fit the spectral data. We observe clear separation between the two sample materials, with water ($\nu_{\rm W}=\Omega_{\rm W}/2\pi$=5.7~GHz) on the left side and plexiglass ($\nu_{\rm P}=\Omega_{\rm W}/2\pi$=11.8~GHz) on the right side. 
The raw non-normalised spectra at all points are shown in Fig.~\ref{fig:plexiglass_spectrum}(b) with the black dashed line showing the mean spectrum calculated across the entire data set. As the scanning plane was chosen just slightly below the plexiglass surface, both water and plexiglass components appear at every measurement point on the plexiglass side of the sample.

We now perform PCA on the full spectral data set to identify the 
main spectral features without using a fitting model. 
If there are $n = N_x N_y N_z$ pixels in the Brillouin measurement volume, each of which yields a spectrum with $m$ frequency bins, then we can represent the data set obtained from Brillouin measurement in the form of a $(n \times m)$ matrix $\mathbf{X}$; the rows of 
$\mathbf{X}$ then correspond to different observations
(which represent different physical positions within the sample), and the
columns correspond to the variables being observed (which represent the different frequencies of spectrum). The central 
idea of PCA is to use a linear transformation $\mathbf P$
to change the data set to a new $(n \times m)$ matrix
\begin{equation}
{\mathbf Y} = {\mathbf X} \mathbf{P} ~,
\label{eq:Ptransform}
\end{equation}
which is better at capturing the full variance of the data
---specifically, $\mathbf{P}$ should be chosen to diagonalise 
the $(m \times m)$ covariance matrix of
${\mathbf Y}$
\begin{equation}
{\mathbf S}_{\mathbf Y} = \frac{1}{n-1} {\mathbf Y}^T{\mathbf Y}~.
\label{eq:covariance}
\end{equation}
This occurs when the 
columns of ${\mathbf P}$ are equal to the eigenvalues of ${\mathbf S}_{\mathbf Y}$. Such a diagonalisation is usually performed using Singular Value Decomposition, but of course in-built packages for performing PCA are available on all the major programming platforms.

The columns of the coordinate transformation ${\mathbf P}$ are known as
the principal components (PCs). The PCs represent the directions in an $m$ dimensional space in which the variance of the data is maximized (see Fig. \ref{fig:plexiglass_spectrum}(c)), and so each principal component can be thought of as representing a particular piece of information associated with the measurement. Each PC of Brillouin data is a vector of length $m$ and so is, effectively, a function of frequency. 
It is therefore tempting to associate each principal component with a particular material in the measured sample. However, this can lead to confusing results, especially since the resulting ``spectra'' can have multiple peaks, and even be negative. Instead, it should be noted that the principal components represent departures from the {\em mean} spectrum as measured over all observations.
The measured spectrum for the $i^{\rm th}$ observation is given in terms of 
the PCs by
\begin{equation}
{\mathbf y}_i = \overline{\mathbf y}_i + \sum_{j=1}^m c_{ij} {\mathbf p}_j.
\label{eq:pcexpansion}
\end{equation}
where the vector constants $c_{ij}$ represent the {\em scores} 
(sometimes referred to as {\em abundancies}) of the $j^{\rm th}$ PC for the $i^{\rm th}$ observation. The sum in (\ref{eq:pcexpansion}) is
usually truncated to include the smallest number of PCs that are able to explain the variance.

According to (\ref{eq:pcexpansion}), each data point can be represented as a combination of PCs, which in turn represent the direction of greatest variance in the data. This is graphically shown in Fig.~\ref{fig:plexiglass_spectrum}(c), which shows the 
decomposition of the measured spectra at two specific frequencies into the first two principal components. Here the black vectors represent the principal components $\mathbf{p}_1$ and $\mathbf{p}_2$ , with the scores of the j-th sample point for these first two PCs shown in red.

\begin{figure}[H]
\centering\includegraphics[width=\textwidth]{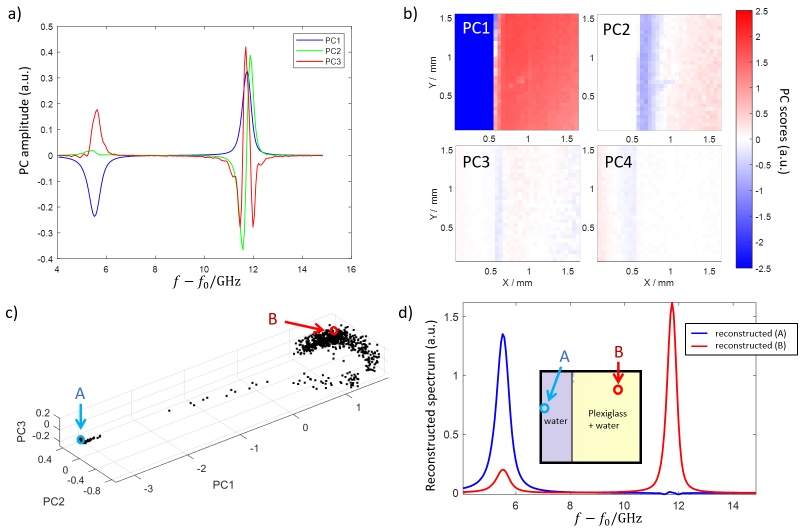}
\caption{Principal components computed from Brillouin microscopy data from the plexiglass sample (see Fig.1a).
a) Spectral amplitude of the first three PCs. The Brillouin peaks are associated with strong variations in each of the PCs. b) Mapping across the sample of the scores of the first four PCs. c) The scores of the first three PCs plotted against each other. Similar materials, giving rise to similar spectra, can be seen to cluster together. The point A is located entirely within the water layer, while the point B is located within the plexiglass/water layer. d) Reconstructed spectra at the points A and B using the first three PCs. The reconstruction fits the measured spectrum at these data points to visual accuracy.
}
\label{fig:plexiglass_PCs}
\end{figure}

The PC amplitudes (for the first three components) for the water/plexiglass sample are shown in Fig.~\ref{fig:plexiglass_PCs}(a). 
We note that these spectral functions 
are multi-peaked and negative - this is a consequence of the fact that they represent
 variations in the data with respect to the mean measured spectral values. 
The scores for the first four PCs are shown
as a map across the sample region in Fig.~\ref{fig:plexiglass_PCs}(b):
here each score represents the amplitude of the given PC at the corresponding spatial location. We use a red-blue colour scheme for which white represents a score of zero. 
The difference between the regions, as well as the edge of the water/plexiglass layer, is visible in the first four components. It can be seen that the PC scores
decrease (become more “white”) for higher order PCs.

\subsection{Extraction of spectra and identification of materials}
The main goal of PCA is to identify the spectra of different materials in the sample region. While the PCs contain information that describes the greatest variation in the data, it is important to keep in mind that this variation may not correlate with the underlying spectrum. For example, even after normalisation, 
the measurement may have a uniform drift in amplitude across the sample volume. Another problem that frequently occurs in 
Brillouin microscopy is that measurements of two different materials 
(say, water and cell cytoplasm) may occur within the same observation;
this makes the distinction between the spectra challenging.

We can extract the underlying material spectra by noting that 
similar materials should possess similar PC scores because the spectrum of each constituent material is the same.
We then expect that observations of the same material 
will cluster together in the $m$-dimensional space of PCA scores. If only two (or three) PCs are needed to explain the data, then the observations of a given material will cluster together in two (or three) dimensions. We observe this clustering in Fig.~\ref{fig:plexiglass_PCs}(c).  From the spatial distribution of data points we observe that the majority of points aggregate near the region labeled with B, with another cluster formed at the opposite end of the PC-space around point A. 
If we then examine the spatial locations of points A and B, 
we find that A is located entirely inside water whereas the B is within the plexiglass. We can confirm this by reconstructing the spectra at points A and B using equation (3); the results are shown in Fig.~\ref{fig:plexiglass_PCs}(d), and show a peak near 5.7 GHz for the point A and a high-amplitude peak near 11.8 GHz for point B. We note also that the reconstructed spectra are 
extremely close to the measured spectra, demonstrating that the first 3 PCs are sufficient to explain the measured data to visual accuracy.

\subsection{Clustering and unsupervised separation of materials}
Proceeding from the observation that clusters in PC space correspond to different materials, we now seek an unsupervised method for separating these clusters and thus identifying regions containing different materials in the sample. A straightforward method to achieve this is k-means clustering, in which the clusters are classified according to 
the squared Euclidean distance to the nearest mean of the cluster group \cite{IKOTUN2023178}. K-means clustering is an unsupervised method
for which the number of clusters must be specified as input to the algorithm. It is therefore straightforward to implement if the number of clusters is known, as in the water/plexiglass sample. However,
without preliminary knowledge of the sample properties, it can be difficult to pinpoint the specific number of clusters that carry physically relevant information. Therefore, to identify the 
number of clusters in the sample we combine the k-means algorithm with
the {\em elbow method} (also known as the {\em knee method}). In this method, the number of clusters is increased until the Within-Cluster-Sum of Squares (WCSS) value reaches an inflection
point, which can be located using the Kneedle algorithm of Satopaa {\em et al. } \cite{Satopaa2011}. The true inflection point will be located above the integer predicted by the Kneedle algorithm, and so we choose the number of clusters to be equal to the next highest integer. We note also that care must be taken in implementing the Kneedle algorithm that a sufficiently high range of clusters is tested: the algorithm will converge to the correct inflection point as the number of number of clusters increases.

\begin{figure}[H]
\centering\includegraphics[width=\textwidth]{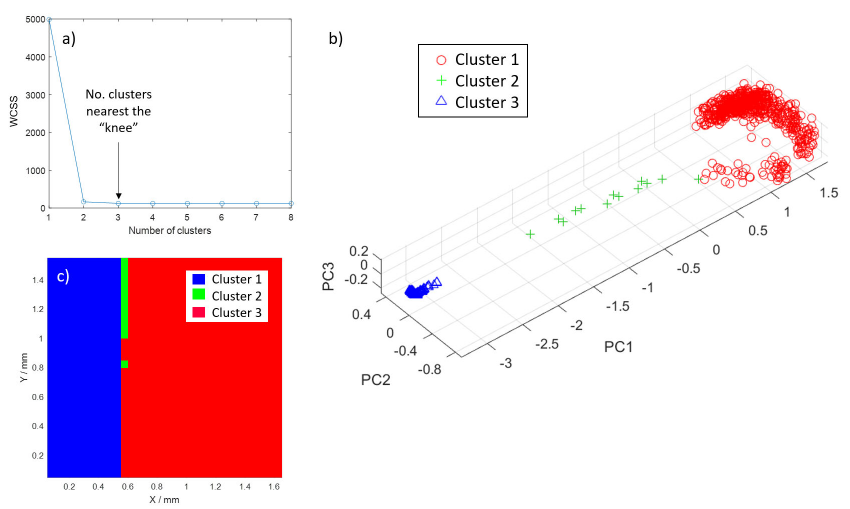}
\caption{Clustering analysis of the PC scores for the water/plexiglass sample. a) Computed Within-Cluster Sum of Squares (WCSS) for different numbers of clusters. The point closest to the ``knee'' is at 3 clusters. b) Plot of PC scores grouped into three clusters, which can be associated with water (blue), plexiglass/water (red) and a boundary region (green). c) Mapping of the three cluster regions across the sample.}
\label{fig:plexiglass_clustering}
\end{figure}

We show the WCSS curve for different numbers of clusters in Fig.~\ref{fig:plexiglass_clustering}(a). 
The position of the ``knee'', as given by the Kneedle algorithm, is between clusters 2 and 3, indicating the optimal number of clusters as 3. In
Fig.~\ref{fig:plexiglass_clustering}(b)) we show the grouped clusters in PC space - one can see the two
clusters corresponding to water (blue) and plexiglass/water (red) have been correctly grouped together by the k-means algorithm, together with a 
sparse collection of points (green) connecting them. 
These points form a third cluster along the interface
between the two layers. We can see this by mapping the 
cluster number of each point to its physical location:
in Fig.~\ref{fig:plexiglass_clustering}(c) one can see
that the different clusters correspond to 
regions of water, plexiglass,
and the thin interface between them.

A different clustering algorithm may divide the points in PC space in a different way: for example, one can see in 
Fig.~\ref{fig:plexiglass_clustering}(b)) a section of the 
plexiglass cluster (red) that is slightly separated from the main cluster, and for which a different clustering algorithm may allocate to a different or its own cluster (for this example these correspond to points near the interface on the plaxiglass side).
To separate these points one could use a more sophisticated algorithm (such as DBSCAN), or resort to supervised division of the clusters \cite{ester1996density}.

\section{Analysis of Brillouin data from a hydrogel spheroid}
\begin{figure}[H]
\centering\includegraphics[width=\textwidth]{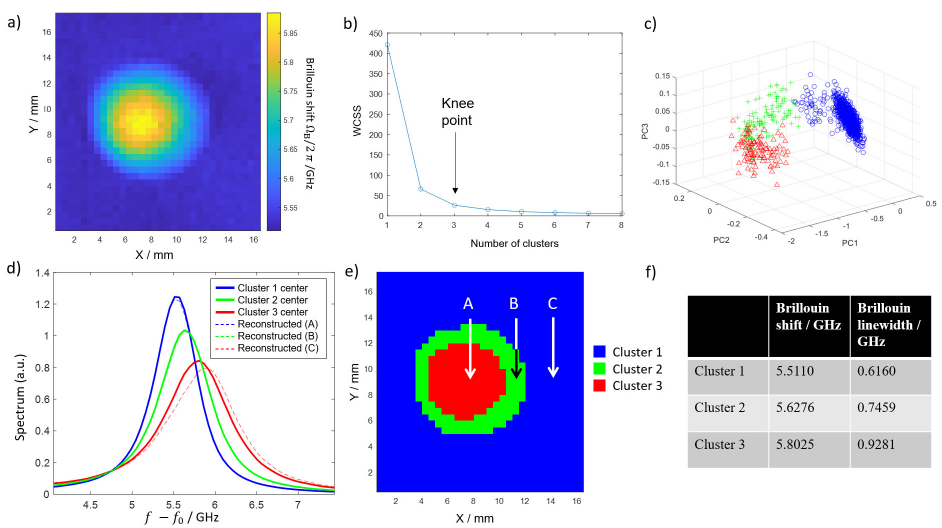}
\caption{PCA and clustering for the hydrogel sample (compare to Fig. 1b). a) Mapping of BFS across the sample. b) Clustering analysis of PC scores, with knee point computed at 3 clusters. c) PC scores for the first three PCs, divided into three clusters. d) Brillouin spectrum at the mean point of each cluster,
compared with the reconstructed spectrum at points at the very centre of the hydrogel sample (A), on the boundary of the sample (B) and within the exterior region (C). Both BFS and linewidth increase toward the centre of the hydrogel. e) Mapping of the cluster groups in the different parts of the sample. The points A,B and C show the points for the spectra in (d). f) Computed BFS and linewidth at the center of each cluster,
using the spectra obtained from the first three PCs.}
\label{fig:hydrogel}
\end{figure}

We now repeat the same procedure for the second sample (see Fig. \ref{fig:fig1_samples}(b)), a hydrogel spheroid. The fitted Brillouin frequency map is shown in Figure~\ref{fig:hydrogel}(a). The WSCC values are shown in 
Fig.~\ref{fig:hydrogel}(b), and identify 3 clusters, which are represented in 
PC space for the first three PCs in Fig.~\ref{fig:hydrogel}(c). Here the clusters are not noticeably distinct, forming a continuous spread in PC space. We hypothesize that the swelling process leads to non-uniform distribution of the mechanical properties, forming a core-shell structure with a stiffer core in the middle of the spheroid and softer, more hydrated shell around it. 

We plot a representative spectrum for each of the three clusters by selecting the central point in each cluster (solid lines in Fig.~\ref{fig:hydrogel}(d)) and compare this to spectral functions reconstructed for the points A, B and C taken within each of the cluster group. We observe good agreement between the reconstructed spectral lines and the spectrum derived at the centre of the corresponding cluster. The identified regions are shown in Fig. \ref{fig:hydrogel}(e), where the three clusters correspond to the interior, boundary and exterior of the spheroid. Finally, we can obtain the Brillouin frequency shift and Brillouin linewidth at the centre of each cluster independently; the results are shown in the table of Fig.~\ref{fig:hydrogel}(f). As expected, the Brillouin frequency shift is the largest in cluster 3, which corresponds to the spheroid core, while reducing slightly for the shell structure at the spheroid's periphery. The linewidth also decreases 
as we move from the interior to the exterior of the spheroid.

\section{Analysis of Brillouin data from a human lung fibroblast cell}
\begin{figure}[H]
\centering\includegraphics[width=\textwidth]{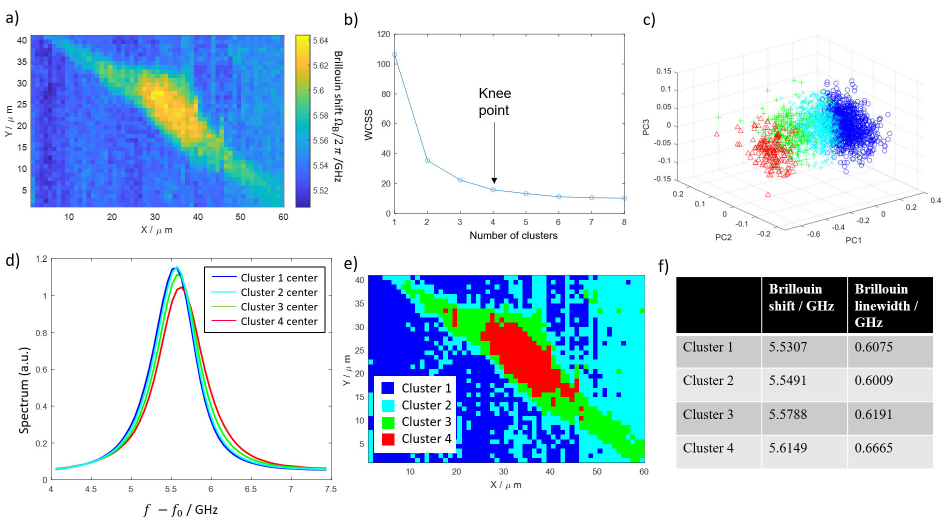}
\caption{PCA and clustering for the human fibroblast cell (compare to Fig. 1c). a) Mapping of BFS across the sample. b) Clustering analysis of PC scores, with knee point computed at 4 clusters. c) PC scores of the first three PCs, divided into four clusters. d) Brillouin spectra computed at the central point of each cluster. c) Mapping of the clusters in the different parts of the sample. e) Computed BFS and linewidth at the centre of each cluster,
using the spectra obtained from the first three PCs.}
\label{fig:cell}
\end{figure}

The spatial distribution of Brillouin frequency shifts in the third sample, an MRC fibrolast cell (see Fig. \ref{fig:fig1_samples}(c)), is depicted in Figure \ref{fig:cell}(a). Higher BFSs are observed in the cell centre compared to the cell periphery. This suggests a larger longitudinal modulus associated with the cell centre, although this can only be confirmed if the distribution of the refractive index and mass density is known or can be measured in parallel with BFS detection \cite{eLife}.  
The WSCC curve (Fig. \ref{fig:cell}(b)), identifies the knee point at four clusters, shown in PC space for the first three PCs in 
Fig. \ref{fig:cell}(c).
As with the hydrogel sample, the PC scores form a continuum of points rather than distinct groups. 
 
In Fig.~\ref{fig:cell}(d), the Brillouin spectra for the centre of each cluster are plotted. These spectra exhibit slight variations in peak positions and intensities, which suggest differences in the mechanical properties between the regions. 
The regions corresponding to the different clusters are shown in Fig.~\ref{fig:cell}(e). The cell centre is clearly visible, as is the
outer layer corresponding to cell cytoplasm and membrane.
The frequency shifts and linewidths of the cluster centres are tabulated in Fig.~\ref{fig:cell}(f). 
These measurements indicate a range in material stiffness and structural heterogeneity across the clusters.
We see that the Brillouin shifts and linewidths of the exterior region 
(Clusters 1 and 2) are very close, and may arise because the clustering approach is identifying differences between the immersion liquid, PBS, and the fibronectin layer on which the cells are seeded, which is much thinner (approx. 3-5 micron) than the axial spatial resolution. 

\section{Discussion}
We now discuss our findings and compare how these align with the previously reported in the literature. First, we note that our main intention for this work was to provide a workflow guide to PCA method in application to Brillouin microscopy data. As mentioned previously, this method is still relatively underutilised with only a few reports available to date \cite{Xiang2021,Palombo_Analyst2018,Cardinali2023}. It is possibly due to this lack of use, many features of the method have not been previously explained.  This could lead to misinterpretations and confusion by the research community, as well as rejection of the method altogether. 

One of the features that have not been understood or interpreted correctly involves the functional form of the principal components. As discussed in Section 3.1, these can become negative-valued and may be confusing to interpret as these PC functions do not always resemble the measured spectra. Once understood that the PC functions need to be viewed as variations from the "mean" spectra and corrected accordingly, the true functional form of principal components can be reconstructed (Figs.~5(d) and 6(d)). Such reconstruction may be particularly valuable for analysing the data collected from Brillouin imaging of heterogeneous materials such as cells. Most of the Brillouin imaging methods suffer from insufficient spectral resolution, with stimulated Brillouin imaging  \cite{Remer} and scanning Fabry-Perot \cite{Alunni} techniques achieving the best resolution to date (approx. 100 MHz). Similarity of the material components in a cell gives rise to the close position of different components' Brillouin peaks in the spectral domain, sometimes within a spectral bandwidth of 100-200 MHz (as in our cell sample) and similar to the instrument's resolution, leading to spectral overlaps between the features. Thus, direct observation and identification of material components often is not possible with the current level of technology. Hence, PCA or other supervised and unsupervised methods might present the only solution to disentangle the complex spectral data collected from heterogeneous samples.

In this work we have presented a new method in which PCA is coupled with k-means clustering to assist with data interpretation. We see this as a complimentary approach, that is particularly valuable in situations where large volumes of data need to be processed without preliminary knowledge of the sample structure and content. In that specific situation, the discretization of all data into clusters that can be analysed separately and compared statistically against different control and treatment groups, can provide the means for rigorous and unbiased analysis. We note however, that for many sample types, including cells and non-uniform biomaterials such as hydrogels as shown in this article, the Brillouin spectral signatures represent a continuum rather than naturally discrete data. This is evident from the PC space for our hydrogel and cell samples (Figs.~\ref{fig:hydrogel}(c) and \ref{fig:cell}(c)). Thus, any division of the data into artificially created bins may lead to errors, especially for data points at the boundary between any two neighboring clusters. The knee (elbow) method is a good tool to assess the cluster number and avoid "over-binning" of the data. However, it does not resolve the more conceptual conflict between the discrete and continuum nature of spectral information and potential risk assigning the wrong labels to a small number of data points. Depending on the data set volumes and the tasks at hand, the risk of error might be fairly insignificant when weighted against the benefits of the clustering method.   

Similarly to the report by Xiang {\it et al.} \cite{Xiang2021}, we identified the value of data pre-processing techniques to enhance the performance of PCA method. Data spectral alignment and normalisation are the main procedures that need to be considered. We confirmed that the quality of the analysis and the distribution of the PC scores are heavily dependent on these pre-processing steps. Surprisingly, we also found that the scores of the principal components were dependent on the data signal to noise ratio (see Supplementary information). This suggests that PCA method might not be the best option for the analysis of low SNR data collected with VIPA-based spectrometers where the imaging speed has been prioritised over the signal quality. Failure to correct for imaging artifacts such as frequency drifts coupled with low SNR data may indeed lead to PCA method producing spurious features devoid of physical meaning, and should be avoided. 

\section{Conclusion}
In conclusion, we discussed the application of principle component analysis of data collected from Brillouin microscopy experiments and provided the guide to data workflow, including pre-processing steps to improve data quality and data transformation into principle components and clusters. For large sets of data (hundreds to thousands of spectra) PCA presents significantly faster method compared to spectral line fitting (fractions of a second compared to minutes). Additionally, PCA does not require any preliminary knowledge of the sample composition, structure, nor guesses of a suitable line shape model. This presents a significant advantage when dealing with complex, heterogeneous samples, for which a single line shape fitting typically results in a significant error. Additionally, we have proposed a combination of PCA with a k-means clustering method that we believe is particularly suitable for biological and biomedical studies with high sample throughput and the need for statistical analysis across various sample groups.

\section{Acknowledgements}
The authors acknowledge the support by the Australian Research Council Centre of Excellence in Optical Microcombs for Breakthrough Science (CE230100006) and the Australian Research Council Centre of Excellence in Quantum Biotechnology (CE230100021).


\bibliography{library}

\pagebreak
\begin{center}
\textbf{\Large Supplemental Materials}
\end{center}

\title{Principle Component Analysis in Application to Brillouin Microscopy Data}

\author{Hadi Mahmodi\authormark{1\dag}, Christopher G. Poulton\authormark{1 \dag}, Mathew N. Lesley\authormark{2},  Glenn Oldham\authormark{3}, Hui Xin Ong\authormark{2}, Steven J. Langford\authormark{1}, and Irina V. Kabakova\authormark{1*} }
\authormark{1} School of Mathematical and Physical Sciences, University of Technology Sydney, NSW, Ultimo, Australia  \newline
\authormark{2} Respiratory Technology, Woolcock Institute of Medical Research, NSW, Glebe, Australia  \newline
\authormark{3} Swinburne University of Technology, Melbourne, Victoria, Australia   \newline
\authormark{*} irina.kabakova@uts.edu.au  \newline
\authormark{\dag} equal contributions \newline

\begin{figure}[H]
\centering\includegraphics[width=10cm]{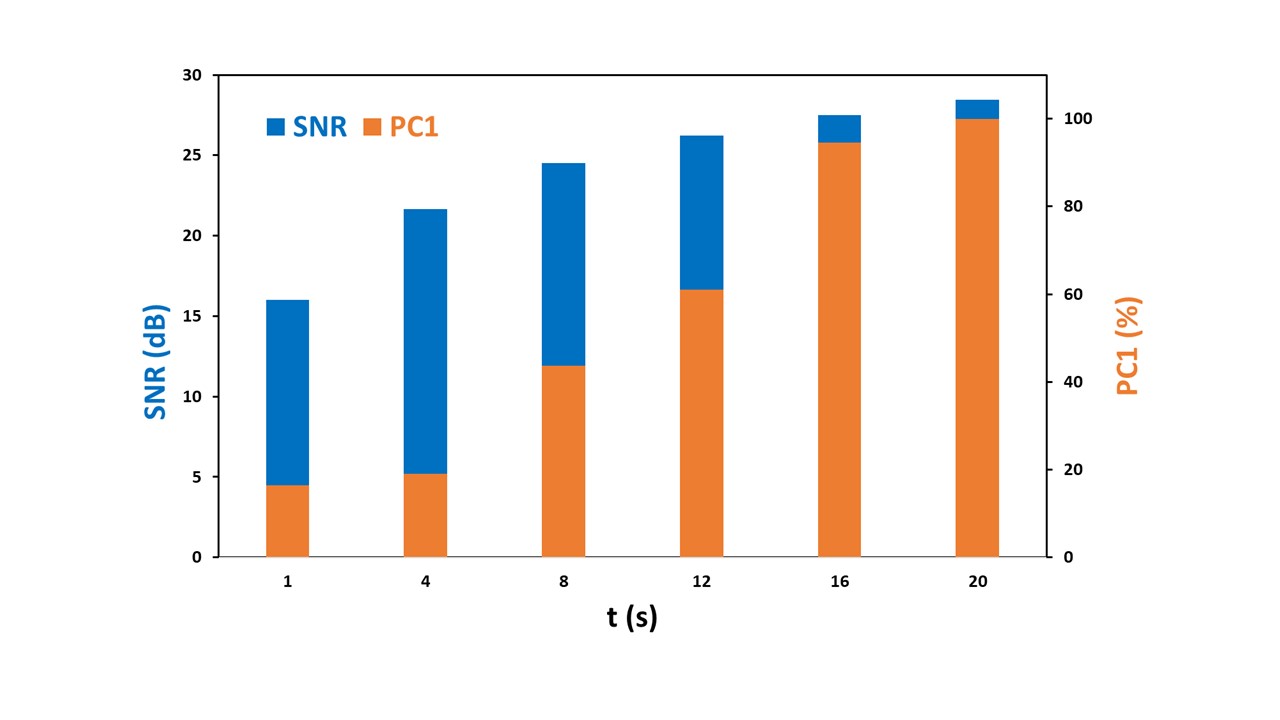}
\break{Fig. S1. SNR and PC1 for Brillouin measurements of distilled water taken over 1-20 s.}
\label{fig:PCA_SNR}
\end{figure}

Figure S1  shows the SNR and percentage of the first (main) Principle Component (PC) in the Brillouin scattering measurements of distilled water (N=25 repeats). Since water is an isotropic and homogeneous liquid, the Brillouin scattering spectrum of water should contain a single peak, and PCA should produce a single component that carries most of information in the data set. We found however, that the percentage of PC1 depends on data’s SNR. For SNR=16, PC1 component represented below 20$\%$ of all data. The percentage grew with increasing SNR value, reaching 99$\%$ for acquisition time of 20 seconds and SNR=27. This highlightes the importance of data quality and suggests that noise reduction techniques, either by applying longer acquisition time or via de-noising methods in post-processing, might be needed to extract reliably features related to sample intrinsic properties rather than fluctuations added by noise.

\begin{figure}[H]
\centering\includegraphics[width=\textwidth]{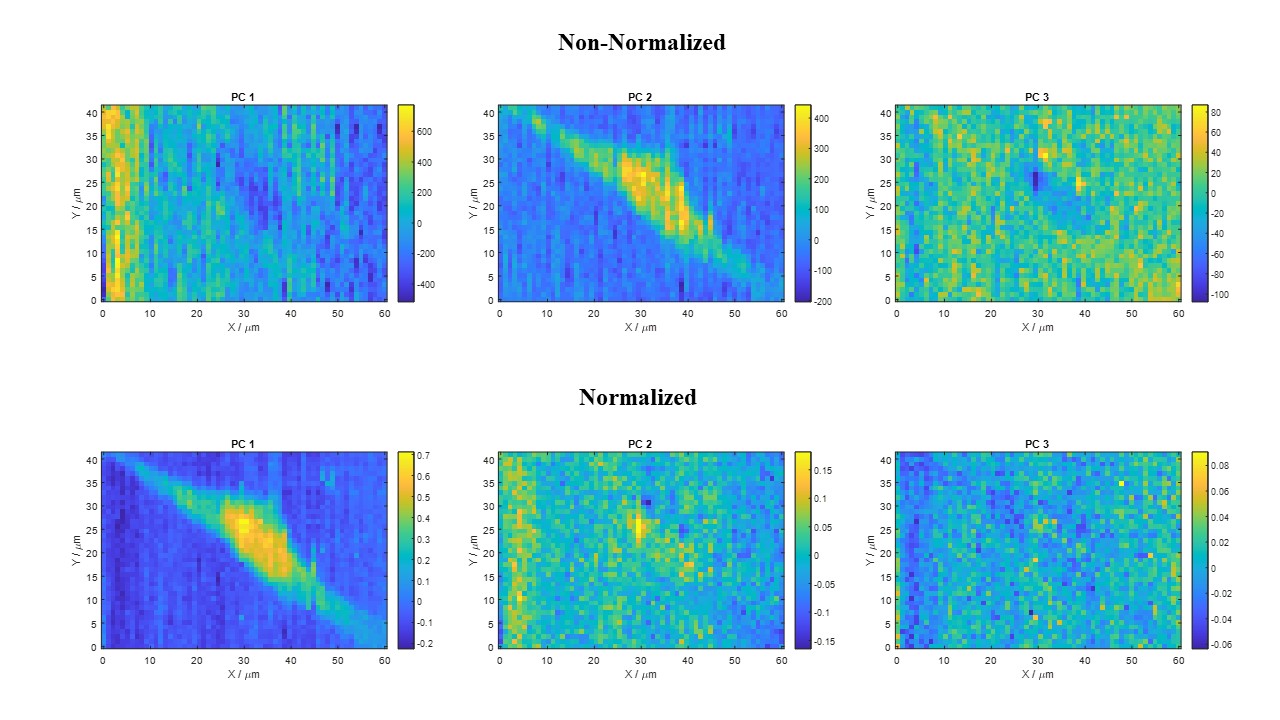}
\break{Fig. S2. The comparison of the first three Principal Components (PCs) in non-normalized vs. normalized data.}
\label{fig:Normalized}
\end{figure}

Figure S2 shows abundance map of first three PCs of non-normalized and normalized cell data. This normalization process enhances the interpretability of the PCs by reducing the influence of background drift and allowing for a clearer identification of patterns and clusters within the cell data. This comparative approach underscores the impact of data processing techniques on the extraction of biologically relevant patterns and the importance of appropriate data normalization in multivariate analysis.

\end{document}